\documentclass[12pt]{iopart}

\usepackage{iopams}  
\begin{document}

\title[Unconventional Color Superconductor]{Unconventional Color Superconductor}

\author{Mei Huang}
\address{Institute of High Energy Physics, Chinese Academy of Sciences, \\
Beijing 100049, China}
\ead{huangm@mail.ihep.ac.cn}

\begin{abstract}
Superfluidity or superconductivity with mismatched Fermi
momenta appears in many systems such as charge neutral 
dense quark matter,  asymmetric nuclear matter, and in 
imbalanced cold atomic gases. The mismatch plays the role of 
breaking the Cooper pairing, and the pair-breaking state cannot be 
properly described in the framework of standard BCS theory.  I give a brief 
review on recent theoretical development  in understanding unconventional 
color superconductivity, including gapless color superconductor, the 
chromomagnetic instabilities and the Higgs instability in the gapless phase.
I also introduce a possible new framework for describing unconventional 
color superconductor.

\end{abstract}


\section{Introduction}

The research of charge neutral cold dense quark matter has driven the color 
superconductivity theory into a new territory beyond standard BCS theory.  The requirement 
of charge neutrality condition induces a substantial mismatch between the 
Fermi surfaces of the pairing quarks, which reduces the available phase space for Cooper pairing, 
and eventually breaks the Cooper-pairing state. Superfluidity or superconductivity with mismatched 
Fermi momenta also appears in other systems, e.g., electronic superconductor in a strong external 
magnetic field,  asymmetric nuclear matter, and in imbalanced cold atomic systems. 
However, it has been an unsolved old problem how a BCS superconductivity is destroyed 
as the mismatch is increased. 

 The results based on conventional BCS framework brings us puzzles. 
For example, in the standard BCS framework, it was found that at moderate mismatch, 
homogeneous gapless superconducting phases \cite{g2SC,gCFL} can be formed. 
However,  gapless superconducting phases exhibit anti-Meissner screening 
effect, i.e., chromomagnetic instability \cite{chromo-ins-g2SC,chromo-ins-gCFL}, 
which is in contradict with the Meissner effect in standard BCS superconductivity. 

I give a brief report on recent theoretical development  of understanding the puzzles 
in g2SC phase.  Firstly, I explain that the gapless 
phase is the result under the conventional BCS framework in mean-field approximation, 
then I introduce a possible proper new framework.  
 
 \section{Mismatch induced gapless phase in mean field approximation}
 \label{sec-gapless}

In the mean-field approximation, the order parameter of 2SC phase takes a constant
and has the form of $\Delta (x) = |\Delta(x)|=\Delta$, and the free energy  for $u, d$ quarks in 
$\beta$-equilibrium  has the form \cite{g2SC}:
\begin{eqnarray} 
\Omega_M = 
 \frac{\Delta^2}{4G_D}
-\sum_{A} \int\frac{d^3 p}{(2\pi)^3} \left[E_{A}
+2 T\ln\left(1+e^{-E_{A}/T}\right)\right] .
\label{pot-m}
\end{eqnarray} 
The sum over $A$ runs over all (6 quark and 6 antiquark)
quasi-particles. 

With the increase of mismatch, the ground state
will be in the gapless 2SC phase when $\Delta < \delta\mu$. 
The gapless phase is in principle a metastable Sarma state \cite{Sarma}, i.e., the free energy
is a local maximum with respect to the gap parameter $\Delta$, and one has
\begin{equation}
\Big(\frac{\partial^2\Omega_M}{\partial\Delta^2}\Big)_{\bar\mu,\delta\mu}
=\frac{4\bar\mu^2}{\pi^2}\Big(1-\frac{\delta\mu}{\sqrt{\delta\mu^2-\Delta^2}}\Big) < 0
\label{sarmains}
\end{equation}
in the gapless phase.

As we already knew that g2SC phase exhibits anti-Meissner screening 
effect or chromomagnetic instability \cite{chromo-ins-g2SC}, 
which is in contradict with the Meissner effect in standard BCS superconductivity. 

\section{The new theoretical framework}

In recent years, there have been lots of effort trying to understand the puzzle of anti-Meissner 
effect, here I give a brief report from my own point of view.   
  
In the 2SC phase,  the color symmetry $G=SU(3)_c$  breaks to subgroup $H=SU(2)_c$.
The full order parameter of the 2SC phase is characterized by
\begin{equation}
\Delta(x) =
\exp \left[ i \left( \sum_{a=4}^8 \varphi_a(x) T_a \right) \right]  (0, \, 0, \, \Delta + H(x)),
\label{II}
\end{equation}
where $\varphi_a (a=4,\cdots,8)$ are five Nambu-Goldstone 
diquarks describing the phase fluctuation of the order parameter, 
and $H(x)$ is the Higgs field describing the spatial fluctuation of the order parameter.

Expanding around the ground state: $(0, \, 0, \, \Delta)$,
the free energy of the system takes the expression as
$\Omega = \Omega_M +\Omega_{NG} + \Omega_{H}$,
where $\Omega_M$ is the contribution from the mean-field approximation
and was given in Eq. (\ref{pot-m}), $\Omega_{NG}$ and $\Omega_{H}$
are contributions from the fluctuation of Nambu-Goldstone currents and
Higgs field, respectively.

 \subsection{Nambu-Goldsone current generation and single plane-wave state}
 \label{sec-current}
 
 The quadratic action of the Goldstone modes in the long wavelength limit can be written down 
with the aid of the Meissner masses $m_a^2$ evaluated in Ref.\cite{chromo-ins-g2SC}, it takes the form of
\begin{equation}
\label{pot-NG}
\Omega_{NG}=\frac{1}{2}\int d^3 {\vec r} \sum_{a=1}^8 m_a^2 \left[ {\vec {\bf A}}^a 
- \frac{1}{g} {\vec \triangledown}  \varphi^a \right] \left[ {\vec {\bf A}}^a- \frac{1}{g}
{\vec \triangledown}  \varphi^a \right] 
+ higher \, orders \, .
\end{equation}
It was found that at
zero temperature, with the increase of mismatch,  for five gluons with $a=4,5,6,7,8$ 
corresponding to broken generator of $SU(3)_c$, their Meissner screening mass squares
become negative \cite{chromo-ins-g2SC}. This indicates the spontaneous generation of 
Nambu-Goldstone currents $\sum_{a=4}^{ 8}  <{\vec \triangledown}  \varphi^a> \neq 0$ \cite{
NG-current-hong, NG-current-huang, NG-current-gCFL} or 
gluon condensate state $\sum_{a=4}^{ 8} {\vec {\bf A}}^a \neq 0$ \cite{GHM-gluon}. 
The NG current state can also be interpreted as
a colored Larkin-Ovchinnikov-Fulde-Ferrel (LOFF) state with the single plane-wave order parameter 
$\Delta(x) = \Delta {\rm exp} ( i \sum_{a=4}^{8}  {\vec \triangledown}  \varphi^a \cdot {\vec {\bf x}})$.

The LOFF state \cite{LOFF-orig} was proposed in 1960s to describe a possible ground state of the 
pair-breaking state of an electronic superconductor when applying a strong external magnetic field. 
However, it has still not yet been confirmed experimentally in electronic 
superconduting systems, and it still remains being persued after 
more than 40 years. Imbalanced cold atom systems offer another intriguing 
experimental possibility to understand how Cooper pairing is destroyed. Due to 
the absence of both the orbital effects and impurities, it seems very promising to search 
for the LOFF state in imbalanced cold atom systems. 
However, recent experiments \cite{LOFF-Exp-atom} in imbalanced cold atom 
systems did not show evidence of the formation of the LOFF state,  rather indicated a 
non-uniform state of phase separation state. 

Obviously, there is something missing in our understanding of the pairing breaking state
if we failed to observe the (LO)FF state in the mismatch regime where the magnetic or superfluid 
density instability develops.

\subsection{Higgs/amplitude  instability and spatial inhomogeneity}
\label{sec-higgs}

The free energy from the Higgs field can be evaluated and takes the form of
\begin{eqnarray} 
\label{pot-H}
\Omega_{H} & =&  \frac{T}{2}\sum_{n}\int\frac{d^3\vec{k}}{(2\pi)^3} 
 H^*(\vec k)\Pi_{H}(k) H(\vec k).
\end{eqnarray}
It was found in Ref. \cite{GHHR} that the self-energy of the Higgs field, $\Pi_H(k)$,
becomes negative in the gapless phase when $\delta\mu > \Delta$, the same type
of  instability was also discussed in Refs. \cite{NG-Hashimoto, Kei-Kenji}.
 
Negative $\Pi_H(k)$ indicates the Higgs mode is unstable 
and will decay. The numerical results in Ref. \cite{GHHR} shows that $\Pi_H(k)$ reaches 
its minimum at a momentum, i.e., $k \simeq 4 \Delta$, which indicates that a stable state may develop
around this minimum, we characterize this momentum as $k_{min}$. 
The inverse $k_{min}^{-1}$ is the typical wavelength for the unstable mode.
If mixed phase can be formed, the typical size $l$ of the 2SC bubbles 
should be as great as $k_{min}^{-1}$, i.e., $l \simeq k_{min}^{-1}$, which 
turns out to be comparable to the coherence length of 2SC,
$\xi_0\simeq\Delta_0^{-1}$.
Considering that the coherence length $\xi$ of a superconductor is proportianl to 
the inverse of the gap magnitude, i.e., $\xi \simeq \Delta^{-1}$,
therefore, a rather large ratio of $k_{min}/\Delta$ means
a rather small ratio of $l/\xi$. When $l/\xi <1$, a phase separation state is 
more favorable. 

In the system of imbalanced neutral atoms, the Higgs instability persists and induces
spatial non-uniform phase separation state. This explains why imbalanced cold atom experiments 
did not observe LOFF state rather showed strong evidence of phase separation. 
For the 2-flavor quark matter being considered,  it was found in Ref. \cite{GHHR}
that,  the electric Coulomb interaction is not strong enough to compete with the Higgs instability.

\section{Summary}

I reviewed the recent theoretical development for describing unconventional color superconductor.
This review is based on my own understanding, and I apologize if I missed some important
work in this field. From my own point of view, I think with the increase of mismatch, it is very essential 
to consider the contribution from the phase fluctuation and the amplitude fluctuation of the order parameter.
The instability from the phase part of the order parameter induces the Nambu-Goldstone currents
generation and forms the single plane-wave state, and the instability from the amplitude part indues the
spatial inhomogeneity and forms the mixed state. The true ground state of a mismatched pairing state
should be determined by the competition of the instabilities. 

\ack
I thank  W. Q. Chao, I. Giannakis, D.F.  Hou, H.C. Ren,  I. Shovkovy, and P.F. Zhuang for 
collaboration.  The work is supported by the Institute of High Energy Physics, Chinese 
Academy of  Sciences(CAS), and CAS key project  under grant  No. KJCX3-SYW-N2.

\end{document}